\documentclass[apl,twocolumn,floats,showpacs]{revtex4}%
\usepackage{graphicx,epsfig}

\begin{document}


\title{Undoped Electron-Hole Bilayers in a GaAs/AlGaAs Double Quantum Well}

\author{J. A. Seamons}
 \email{jaseamo@sandia.gov}
\author{D. R. Tibbetts}
\author{J. L. Reno}
\author{M. P. Lilly}
\affiliation{Sandia National Laboratories, Albuquerque, NM 87185}

\date{\today}

\begin{abstract}
We present the fabrication details of completely undoped electron-hole bilayer devices in a GaAs/AlGaAs double quantum well heterostructure with a 30
nm barrier.  These devices have independently tunable densities of the two-dimensional electron gas and two-dimensional hole gas. We report
four-terminal transport measurements of the independently contacted electron and hole layers with balanced densities from $1.2 \times 10^{11}$
cm$^{-2}$ down to $4 \times 10^{10}$ cm$^{-2}$ at $T = 300 mK$.  The mobilities can exceed $1 \times 10^{6}$ cm$^{2}$ V$^{-1}$ s$^{-1}$ for electrons
and $4 \times 10^{5}$ cm$^{2}$ V$^{-1}$ s$^{-1}$ for holes.
\end{abstract}

\maketitle

We present the details of fabrication and transport of electron-hole bilayers in completely undoped double quantum wells. This work is motivated by
the intense interest in exciton condensation effects that are predicted to occur in electron-hole bilayer systems.  Exciton condensation is theorized
to occur at low carrier densities and close proximity of the two-dimensional electron gas (2DEG) to the two-dimensional hole gas (2DHG). Sivan et
al.\cite{Sivan1992} created an electron-hole bilayer device with the capability to measure Coulomb drag between the layers. However, at low
temperature and low density their ability to probe the exciton condensate was limited. Since then several research groups have studied electron-hole
bilayers using doped
heterostructures.\cite{Shapira1996,Roslund2000,Parlangeli2001,Shapira1999,Keogh2005,Kane1994,Pohlt2002,Sazio1999,Vijendran1999,Parlangeli1998,Rubel1998}
Single layer undoped heterostructures commonly referred to as heterostructure insulated-gate field-effect transistors (HIGFETs) have been
investigated previously.\cite{Lilly2003,Zhu2003,Noh2003}  In these studies particular interest was paid to the ultra-low density capability afforded
by HIGFETs, and the ability to directly control the density and polarity of the carriers. In an effort to utilize these unique abilities of HIGFETs
for the studies of exciton condensation we developed a fully undoped electron-hole bilayer (uEHBL) device.

Significant progress has been made in the areas of two-dimensional bilayer systems with regard to electron-electron or hole-hole bilayers. Transport
exciton condensate experiments at a zero magnetic field in electron-hole bilayers have proved to be extremely difficult. Some of the previously
explored methods failed to make independent contact to the electron and hole layers (including photoluminescence studies) or required a large barrier
thickness.\cite{Rubel1998}  Most of the previous studies were limited by an inability to adjust the densities in the two layers sufficient to match
up the density of the 2DEG ($n$) to the density of the 2DHG ($p$).  This was typically either due to the use of a doped
heterostructure\cite{Roslund2000,Parlangeli2001} or a design which lacked a back gate.\cite{Shapira1999,Keogh2005,Drndic1997}  The uEHBL architecture
allows for independent contacts to each layer, high mobility, and tunable low densities for the 2DEG and 2DHG.  Therefore the advantage of uEHBLs is
that exciton condensation is potentially achievable with this architecture.


In this letter, we describe the fabrication, operation, and limitations of uEHBLs.  We demonstrate the high transport mobility in these samples over
a wide range of electron and hole densities.  Then we examine the magnetoresistance of the 2DEG and 2DHG, and discuss the screening effects that
occur. Finally, we show that $n$ and $p$ can be balanced over a wide range and, more specifically, at the lowest matched densities to date.

\begin{figure}[t]
\centerline{\epsfig{file=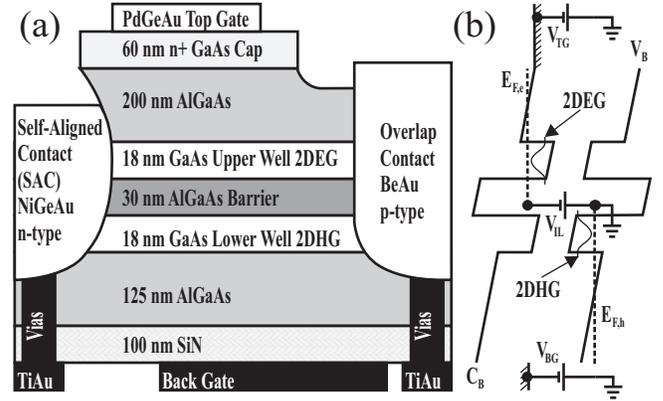,width=8.5cm}} \caption{(a) Cut-away schematic of the device structure. (b) Band diagram schematic with V$_{TG}$,
V$_{IL}$, and V$_{BG}$ sufficient to form the 2DEG and 2DHG.} \label{Figure 1}
\end{figure}

The fabrication of a uEHBL device begins with a molecular beam epitaxy (MBE) grown GaAs/AlGaAs undoped double quantum well (wafer EA1286). As is seen
in Fig. 1(a), from the surface there is a 60 nm n+ GaAs cap layer, a 200 nm Al$_{0.3}$Ga$_{0.7}$As layer, an upper 18 nm GaAs quantum well, a 30 nm
Al$_{0.9}$Ga$_{0.1}$As barrier, a lower 18 nm GaAs quantum well, a 125 nm Al$_{0.3}$Ga$_{0.7}$As layer, a superlattice, and finally an
Al$_{0.55}$Ga$_{0.45}$As stop-etch layer.  The only intentional dopants are in the n+ GaAs cap layer and they do not populate the upper quantum well.
A modified self-aligned contact (SAC), originally pioneered by Kane et al.,\cite{Kane1994,Kane1995} is used for the n-type contacts. This design was
extended upon by processing both sides of the device using the Epoxy-Bond-And-Stop-Etch (EBASE)\cite{Weckwerth1996} thinning technique.  The backside
processing is used to create the back gate for the overlap p-type contacts. The combination of these techniques result in devices that have
independent contacts to the 2DEG and 2DHG, and tunable $n$ and $p$.

Processing of an uEHBL is as follows.  The cleaved sample is etched in phosphoric acid to create a Hall-bar with ten contact arms (five n-type, five
p-type).  Only in the central region of the Hall-bar will the 2DEG and 2DHG be directly above one another. Using phosphoric acid the n+ GaAs top gate
is also etched off the regions that will become p-type contacts to prevent shorting.  A shallow n-type ohmic contact (so as to not short the top gate
to the 2DEG) of PdGeAu is evaporated onto one end of the Hall-bar to contact the n+ GaAs top gate. BeAu is evaporated onto the ends of half the
Hall-bar arms to create the p-type ohmic contacts. The device is annealed in a rapid thermal annealer at 475 C for the p-type contacts.  The n-type
ohmic contacts are SACs of NiGeAu which are placed on the remaining Hall-bar arms.  The device is annealed again, at 420 C for the n-type ohmics to
contact the 2DEG.  Once topside processing of the device is finished, the sample is put through the EBASE process.\cite{Weckwerth1996} The topside of
the device is epoxied to a host GaAs substrate.  Then the original substrate is thinned to the stop-etch layer. The stop-etch layer is removed with
hydrofluoric acid.  A SiN film is deposited to reduce the bottom gate leakage current.  Vias are etched through to contact all of the topside
processing. The uEHBL device is finished with a bottom gate of TiAu down the length of the Hall-bar and out five arms to make overlap contacts with
the p-type metal.

Unlike traditional doped heterostructures in which the carriers are inherently present, uEHBLs require unique operation.  Only under specific bias
conditions are the 2DEG and 2DHG simultaneously occupied.  The most important component of an uEHBL is the core HIGFET.\cite{Kane1993,Lilly2003} In
HIGFETs, ohmic contacts are self-aligned to the top gate.  The carriers are drawn into the 2D layer by applying a voltage between the top gate
(V$_{TG}$) and the SACs. There is a voltage threshold (V$_{th}$) that must be reached to create the 2D layer. Once present the density of the
carriers is linearly proportional to $\mid$ V$_{TG} -$ V$_{th}\mid$.

Operation of uEHBLs is more complicated than HIGFETs.  The upper quantum well, the top gate, and the n-type ohmic contacts operate the same as in
HIGFETs.  The n-type ohmics form SAC with the top gate [see Fig. 1(a)].  It can be operated in a single layer mode where the top gate is positively
biased with respect to the SAC.  As in the case of HIGFETs, when V$_{TG} =$ V$_{th}^{n}$, then the 2DEG forms. The top gate extends over the n-type
contact [Fig. 1(a)].  This increases the electric field on the n-type contact, thus enabling the electrons to be pulled from the SAC into the upper
quantum well to establish the 2DEG.  The resulting $n$ is linearly proportional to V$_{TG} -$ V$_{th}^{n}$.  For the 2DEG to remain established it is
sufficient for a net positive bias between V$_{TG}$ and the voltage on the n-type ohmics equal or larger than V$_{th}^{n}$. This voltage on the
n-type ohmics is ground in single layer operations and is referred to as an interlayer bias (V$_{IL}$) when the bilayers is established.  The role of
V$_{IL}$ is visible in Fig. 1(b); it effectively separates the Fermi energy (E$_{F}$) between the conduction band (C$_{B}$) in the upper quantum well
and the valance band (V$_{B}$) in the lower quantum well.

To form the 2DHG, the bottom gate sits on the opposite side of the substrate with an insulating layer between it and the p-type contact.  We refer to
this as an overlap contact, because the bottom gate overlaps the p-type ohmic contact.  It enables the electric field to be such that the holes are
pulled from the p-type contact into the lower quantum well to establish the 2DHG.  If the 2DHG is to be established without the 2DEG, then it is
sufficient for the bottom gate voltage (V$_{BG}$) to be ramped negative until the holes from the p-type contacts create a 2DHG.  The density, $p$, is
linearly proportional to V$_{BG} -$ V$_{th}^{p}$.  To operate the uEHBL device with both the 2DEG and the 2DHG established it is necessary to
compensate for the energy difference, (the bandgap and quantum well offsets) between the upper quantum well conduction band and the lower quantum
well valence band. To do this V$_{IL}\sim~-1.5V$ is applied to the n-type contacts [see Fig. 1(b)].  Unlike the HIGFET, $n$ and $p$ are functions of
all three voltages in the device. This gives an overdefined system and therefore we can match the densities at slightly different values of V$_{IL}$.

The uEHBL devices have limitations set by the ranges of V$_{TG}$, V$_{IL}$, and V$_{BG}$.  As expected V$_{IL}$ is roughly the size of the band gap
in GaAs of 1.5 V.  An interlayer capacitance measurement shows that overlap first occurs when V$_{IL}\simeq -1.37$ V. And by -1.5 V current begins to
flow between the layers. The smallest values of $n$ and $p$ are determined by the uniformity of V$_{th}^{n}$ and V$_{th}^{p}$ for the contacts. The
largest value of $n$ depends upon the current leaking between the SAC and top gate.  And the largest achievable $p$ is dependant upon the V$_{BG}$
when the back gate begins to leak. These factors vary from sample to sample, making them difficult to control.

The transport data that follows is from two uEHBL devices, A and B.  In device A low density was achieved in both the 2DEG and 2DHG, but the 2DEG
only has the Hall resistance capability. In device B all of the n-type and p-type contacts were functional; however, it is limited by larger than
typical V$_{th}^{n}$ values ($n\geq 3 \times 10^{11}$ cm$^{-2}$).

\begin{figure}[t]
\centerline{\epsfig{file=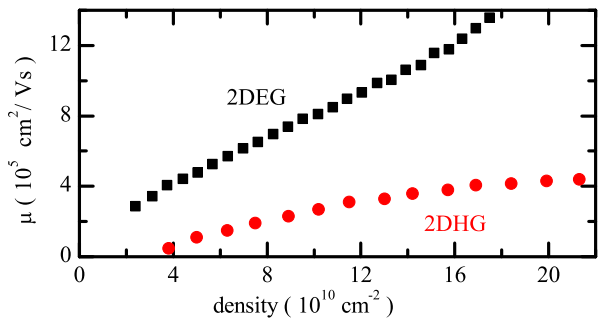,width=8.5cm}} \caption{(Color online) Mobility vs. density, $n$ (black squares) for device A and $p$ (red
circles) for device B} \label{Figure 2}
\end{figure}

The measurements were all taken at $T =300~mK$, using a standard low frequency AC lock-in technique.  In Fig. 2 we present the density versus
mobility for the 2DEG of device A and 2DHG of device B.  The mobilities of the 2DEG and 2DHG may exceed $1 \times 10^{6}$ cm$^{2}$ V$^{-1}$ s$^{-1}$
and $4 \times 10^{5}$ cm$^{2}$ V$^{-1}$ s$^{-1}$, respectively.  The $n$ were calculated from the Hall slope, while $p$ was determined from the
Shubnikov de Haas oscillations.  As expected with ultra-clean 2DEGs and 2DHGs the mobilities increase with densities.

\begin{figure}[t]
\centerline{\epsfig{file=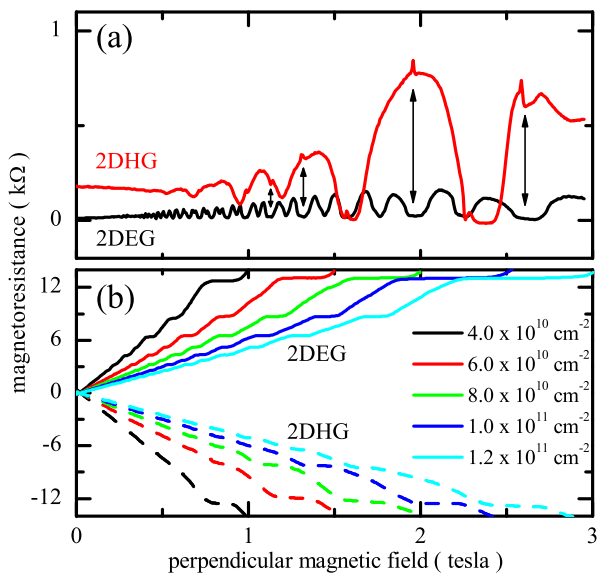,width=8.5cm}} \caption{(Color online) (a) Four-terminal longitudinal resistance of the 2DHG (larger, red), and
the 2DEG (smaller, black) of device B. (b) Matched density Hall resistance traces for the 2DEG (solid) and 2DHG (dashed) of sample A.} \label{Figure
3}
\end{figure}

Magnetoresistance data is presented in Fig. 3.  Fig. 3(a) is the four-terminal longitudinal magnetoresistance with a perpendicular applied magnetic
field from device B with V$_{TG} = +0.07$ V, V$_{IL}= -1.43$ V, and V$_{BG}= -2.0$ V.  The oscillations are more clearly defined for the 2DEG due to
the higher density and mobility.  Clearly a 2DEG and 2DHG were simultaneously established in this device.  When the 2DEG is in a strong quantum Hall
state there is a spike in the 2DHG magnetoresistance [see the arrows in Fig 3(a)].  Since quantum Hall states are incompressible, the effective
screening of the top gate by the 2DEG at that magnetic field changes.  This in turn changes $p$ evident by the spike in the longitudinal
magnetoresistance. In Fig. 3(b) the Hall magnetoresistance traces of device A at V$_{IL}= -1.44$ V in $2 \times 10^{10}$ cm$^{-2}$ increments are
presented. By varying V$_{TG}$ from -0.983 V to -0.733 V and V$_{BG}$ from -1.517 V to -2.02 V the $n$ and $p$ are matched from $4 \times 10^{10}$
cm$^{-2}$ to $1.2 \times 10^{11}$ cm$^{-2}$, respectively.  The opposite slopes of the Hall traces dramatically demonstrate the presence of electrons
and holes.

We report that no evidence for exciton condensation was observed in these uEHBL devices down to the matched $n$ and $p$ of $4 \times 10^{10}$
cm$^{-2}$ with a 30 nm barrier.  At this density the intra-layer spacing is about 50 nm, and the interlayer spacing is 48 nm.  We are presently
fabricating similar devices with narrower barriers.

In conclusion, the uEHBL architecture is ideal for experiments requiring low density, high mobility with layer spacing comparable to average carrier
separation. The operation and fabrication of uEHBLs was described.  High transport mobility over a wide range of $n$ and $p$ was presented.
Capacitance measurements verified the overlap of the 2DEG and 2DHG.  Screening effects as the 2DEG goes through an incompressible quantum Hall state
were shown. Balanced densities as low as $4 \times 10^{10}$ cm$^{-2}$ were achieved.  While exciton condensation has not yet been observed in these
uEHBL devices, continued development holds promise.

We acknowledge outstanding technical assistance from R. Dunn and valuable discussions with  E. Bielejec.  This work has been supported by the
Division of Materials Sciences and Engineering, Office of Basic Energy Sciences, U.S. Department of Energy. Sandia is a multiprogram laboratory
operated by Sandia Corporation, a Lockheed Martin Company, for the United States Department of Energy under Contract No. DE-AC04-94AL85000.


\begin{thebibliography}{}

\bibitem{Sivan1992} U. Sivan, {\em et al.}, Phys. Rev. Lett. {\bf68}, 1196 (1992);
\bibitem{Shapira1996} S. Shapira, {\em et al.}, Phys. Rev. Lett. {\bf77}, 3181, (1996);
\bibitem{Roslund2000} J. H. Roslund, {\em et al.}, Jpn. J. Appl. Phys. {\bf39}, 2448 (2000);
\bibitem{Parlangeli2001} A. Parlangeli, {\em et al.}, Phys. Rev. B {\bf63}, 115307 (2001);
\bibitem{Shapira1999} S. Shapira, {\em et al.}, Appl. Phys. Lett. {\bf74}, 1603 (1999);
\bibitem{Keogh2005} J. A. Keogh, {\em et al.}, Appl. Phys. Lett. {\bf87}, 202104 (2005);
\bibitem{Kane1994} B. E. Kane, {\em et al.}, Appl. Phys. Lett. {\bf65}, 3266 (1994);
\bibitem{Pohlt2002} M. Pohlt, {\em et al.}, Appl. Phys. Lett. {\bf80}, 2105 (2002);
\bibitem{Sazio1999} P. J. A. Sazio, {\em et al.}, J. Cryst. Growth {\bf201/202}, 12 (1999);
\bibitem{Vijendran1999} S. Vijendran, {\em et al.}, J. Vac. Sci. Technol. B {\bf17}, 3226 (1999);
\bibitem{Parlangeli1998} A. Parlangeli, {\em et al.}, Physica B {\bf258}, 531 (1998);
\bibitem{Rubel1998} H. Rubel, {\em et al.}, Mater. Sci. Eng. B {\bf51}, 207 (1998);
\bibitem{Lilly2003} M. P. Lilly, {\em et al.}, Phys. Rev. Lett. {\bf90}, 56806 (2003);
\bibitem{Zhu2003} J. Zhu, {\em et al.}, Phys. Rev. Lett. {\bf90}, 56805 (2003);
\bibitem{Noh2003} H. Noh, {\em et al.}, Phys. Rev. B {\bf68}, 165308 (2003);
\bibitem{Drndic1997} M. Drndic, {\em et al.}, Appl. Phys. Lett. {\bf70}, 481 (1997);
\bibitem{Kane1993} B. E. Kane, {\em et al.}, Appl. Phys. Lett. {\bf63}, 2132 (1993);
\bibitem{Weckwerth1996} M. V. Weckwerth, {\em et al.}, Superlatt. Microstruct. {\bf20}, 561 (1996);
\bibitem{Kane1995} B. E. Kane, {\em et al.}, Appl. Phys. Lett. {\bf67}, 1262 (1995);

\end{thebibliography}
\end{document}